\documentclass[superscriptaddress,amsmath,amssymb,aps]{revtex4-2}

\usepackage{graphicx}
\usepackage{overpic}
\usepackage{xcolor}
\usepackage{dcolumn}
\usepackage{bm}
\usepackage{hyperref}
\usepackage{natbib}
\usepackage{multirow}
\usepackage{siunitx}
\usepackage{tikz}

\definecolor{darkred}{rgb}{0.4,0,0}
\definecolor{darkgreen}{rgb}{0,0.4,0}
\definecolor{darkblue}{rgb}{0,0,0.4}
\definecolor{linegreen}{rgb}{0,0.8,0}

\DeclareMathOperator{\ku}{Ku}

\newcommand{\tauK}{\tau^{(\eta)}}

\newcommand{\gghat}{\hat{\ve g}}
\newcommand{\xxhat}{\hat{\ve x}}
\newcommand{\yyhat}{\hat{\ve y}}
\newcommand{\zzhat}{\hat{\ve z}}
\newcommand{\pzc}{p^{\rm th}_{z}}
\newcommand{\sigmac}{{\sigma^{\rm th}}}
\newcommand{\Deltauc}{{\Delta u^{\rm th}}}
\newcommand{\Sc}{{S^{\rm th}}}
\newcommand{\vvg}{{\ve v^{(\rm g)}}}
\newcommand{\vgp}{v^{(\rm g)}_{\parallel}}
\newcommand{\vgo}{v^{(\rm g)}_{\perp}}
\newcommand{\rhof}{{\rho^{(\rm f)}}}
\newcommand{\rhop}{{\rho^{(\rm p)}}}
\newcommand{\oomegas}{\ve\omega^{({\rm s})}}
\newcommand{\omegasn}{\omega^{({\rm s})}_n}
\newcommand{\omegasp}{\omega^{({\rm s})}_p}
\newcommand{\omegasq}{\omega^{({\rm s})}_q}
\newcommand{\omegas}{\omega^{({\rm s})}}
\newcommand{\vs}{v^{({\rm s})}}
\newcommand{\urms}{u^{\rm rms}}
\newcommand{\Teq}{T^{\rm eq}}
\newcommand{\Ttrain}{T^{\rm train}}
\newcommand{\Tmeas}{T^{\rm meas}}
\newcommand{\Eeps}{E^{(\varepsilon)}}
\newcommand{\Ealpha}{E^{(\alpha)}}

\newcommand{\ve}[1]{\ensuremath{\mbox{\boldmath$#1$}}}
\newcommand{\ma}[1]{\ensuremath{\mathbb{#1}}}
\newcommand{\cross}{\times}

\begin{document}

\title{Active gyrotactic stability of microswimmers using hydromechanical signals}

\author{Jingran Qiu}
\affiliation{AML, Department of Engineering Mechanics, Tsinghua University, 100084 Beijing, China}
\author{Navid Mousavi}
\affiliation{Department of Physics, University of Gothenburg, SE-41296 Gothenburg, Sweden}
\author{Lihao Zhao}
\email{zhaolihao@mail.tsinghua.edu.cn}
\affiliation{AML, Department of Engineering Mechanics, Tsinghua University, 100084 Beijing, China}
\author{Kristian Gustavsson}
\email{kristian.gustafsson@physics.gu.se}
\affiliation{Department of Physics, University of Gothenburg, SE-41296 Gothenburg, Sweden}

\begin{abstract}
Many plankton species undergo daily vertical migration to large depths in the turbulent ocean.
To do this efficiently, the plankton can use a gyrotactic mechanism, aligning them with gravity to swim downwards, or against gravity to swim upwards.
Many species show passive mechanisms for gyrotactic stability. For example, bottom-heavy plankton tend to align upwards.
This is efficient for upward migration in quiescent flows, but it is often sensitive to turbulence which upsets the alignment.
Here we suggest a simple, robust active mechanism for gyrotactic stability, which is only lightly affected by turbulence and allows alignment both along and against gravity.
We use a model for a plankton that swims with a constant speed and can actively steer in response to hydrodynamic signals encountered in simulations of a turbulent flow.
Using reinforcement learning, we identify the optimal steering strategy.
By using its setae to sense its settling velocity transversal to its swimming direction, the swimmer can deduce information about the direction of gravity, allowing it to actively align upwards.
The mechanism leads to a rate of upward migration in a turbulent flow that is of the same order as in quiescent flows, unless the turbulence is very vigorous.
In contrast, passive swimmers with typical parameters of copepods show much smaller upward velocity in turbulence. Settling may even cause them to migrate downwards in vigorous turbulence.
\end{abstract}

\maketitle

\section{Introduction}

Efficient strategies for vertical swimming under turbulence are important for many marine plankton species.
Many plankton species undergo daily vertical migration of up to tens of meters to greater depths and back~\cite{Park2001,Ringelberg2009,Smayda2010} to allow for efficient nutrition uptake, to avoid predators, and to adjust to tidal flows~\cite{Smayda2010,Schmitt2011,Katajisto1998,Hays1994}.
Many species adopt strategies leading to correlated and ballistic vertical migration along or against gravity~\cite{Schuech2014,Pundyak2017}.
Their horizontal dynamics, by contrast, is usually simply diffusive.
There are several simple passive gyrotactic mechanisms to achieve rapid vertical migration. Some organisms are bottom heavy, or display shape asymmetries.
In both cases, the resulting torques anti-align the organism with the direction of gravity.
Many plankton species also use active gyrotactic strategies. Some phytoplankton can adjust their shape to either align with or against gravity~\cite{Sengupta2017}. Species of protists and flagellates sense gravity using mechanosensitive ion channels to allow them to adjust their swimming direction with regards to gravity~\cite{Hemmersbach1999,Haeder2003,HemmersbachKrause1993}.
To our knowledge, there is no evidence that planktonic copepods sense gravity using ion channels.
It has been suggested that planktonic copepods instead can use setae distributed on their body and antennae to obtain information about their orientation relative to gravity~\cite{Strickler1973}, but this hypothesis has not been verified or discussed in detail.
The setae allow copepods to sense slip velocities as small as \SI{20}{\micro\metre\per\s} between themselves and the ambient fluid~\cite{Yen1992}. This is one order of magnitude smaller than their typical settling velocity in quiescent flow~\cite{Titelman2003}.
The setae also allow for measurement of additional hydromechanical signals such as the local strain rate and slip vorticity.
The strain rate is known to be important for predation and predator avoidance~\cite{Kioerboe1999predator}, but the detailed response of the copepod to different hydromechanical signals remains unclear.
Is it possible for copepods to use the information perceived by hydromechanical signals to achieve active gyrotaxis, and if so, what is the best swimming strategy for this?
The answers to these questions may provide a new perspective to understand the vertical migration taking place in different situations~\cite{Schmitt2011,Katajisto1998,Hays1994}.
The answers also have bearing on future applications of fabricated microswimmers.

It is not understood how microswimmers such as copepods navigate in the most efficient way in the turbulent ocean.
Due to their small size and the complexity of the flow, it is difficult to perform experimental investigations on the relationship between hydromechanical signals and navigation strategies such as vertical migration.
To anyway try to understand and explain the dynamics of plankton, a simplified model has been formulated, describing plankton as a point-like, bottom heavy, spheroidal particle that swim with a constant speed relative to the local fluid in their instantaneous direction~\cite{Durham2013,Lillo2014,Zhan2014,Gustavsson2016,Borgnino2018,Lovecchio2019}. Analyzing this model using direct numerical simulations (DNS) of turbulence and analytical approaches allows to discover the mechanisms that determine orientational statistics, clustering, preferential sampling, and vertical migration of plankton.
The gyrotactic torque resulting from the bottom-heaviness of the swimmer aligns it against gravity and give efficient upward vertical migration in the absence of turbulence.
But in the presence of turbulent velocity gradients, the model predicts that the upward vertical-migration velocity of copepods with parameters approximated from nature is only a fraction of the swimming speed, and upward migration may even fail due to gravitational settling~\cite{Qiu2020,Qiu2020a,Qiu2021}.
In addition, the model does not explain how some organisms swim downwards~\cite{Dodson1997,Ringelberg2009}.
Settling is too slow for copepods to migrate several metres to the depths at dawn, they must therefore use an active mechanism to descend more efficiently.

Due to the lack of experimental input, it is hard to formulate improved models where the swimmers can actively adjust their swimming behavior in response to external stimuli.
One approach is to use reinforcement learning to find good strategies, allowing to construct new models that can be compared to experiments.
Proof of concept studies have shown that reinforcement learning provides strategies for efficient vertical migration~\cite{Colabrese2017,Gustavsson2017,Alageshan2020,Qiu2020} and more general navigation tasks~\cite{Alageshan2020,Biferale2019,Schneider2019,Muinos2021,Gunnarson2021} for swimmers with different shapes and motilities in both two- and three-dimensional flows.
However, in these studies the swimmers had access to global information such as their absolute position or orientation.
Usually, the swimmers sense only local information, in their own frame of reference. As a consequence, vertical migration requires that external or hydrodynamic forces break the symmetry of the problem. For example, it was shown in Ref.~\cite{Qiu2021} that the symmetry breaking set by gravity allows the swimmer to infer its vertical orientation by sensing only local hydromechanical signals.
Using reinforcement learning, efficient steering protocols for upward navigation, based primarily on the local strain rate, were found in frozen (time-independent) two-dimensional flows.
The results illustrate that the swimmers can learn rapid vertical migration by mimicking slender swimmers that tend to preferentially sample upwelling regions in the flow~\cite{Gustavsson2016,Borgnino2018,Lovecchio2019,Cencini2019}.

\begin{figure}
\includegraphics[width=15.2cm]{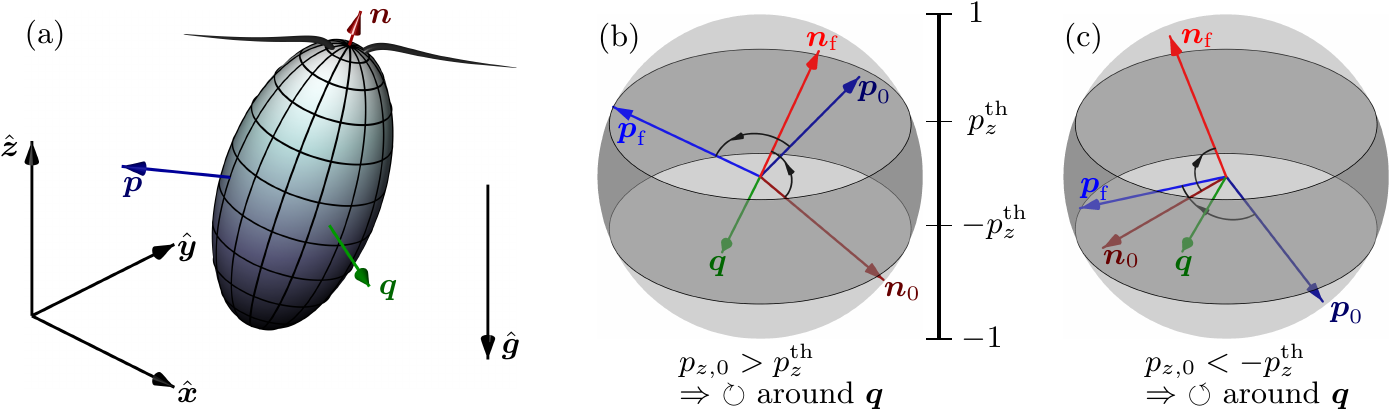}
\caption{\label{fig:mechanism}
(a) Orientation axes $\ve n$, $\ve p$ and $\ve q$ for a spheroidal swimmer in a fixed Cartesian frame of reference ($\xxhat$, $\yyhat$, $\zzhat$) with direction of gravity $\gghat=-\zzhat$.
(b, c) Unit spheres illustrating the mechanism in Eq. (\ref{eq:optimal}) for active gyrotactic stability.
Starting with orientation $\ve n_0$, $\ve p_0$, and $\ve q$, the swimmer actively rotates around its $\ve q$-axis until $|p_z|\le \pzc$ (shaded region) with positive rotation if $p_{z,0}>0$ (b) and negative rotation if $p_{z,0}<0$ (c), leading to the final orientation $\ve n_{\rm f}$, $\ve p_{\rm f}$, and $\ve q$ with $n_{z,{\rm f}}\ge n_{z,0}$ in general.}
\end{figure}

The mechanism mentioned above is specific to two-dimensional flows. In this paper we therefore investigate vertical-migration strategies in three-dimensional flows. We use reinforcement learning to find candidate strategies. The best strategies found are similar to those found in the two-dimensional frozen flow~\cite{Qiu2021}, but it turns out that the slip velocity is the dominant signal, in contrast to the strain rate for the frozen two-dimensional flows.
The main result is a simple yet powerful strategy for vertical migration based on a single component of the slip velocity, summarized in Fig.~\ref{fig:mechanism}.
Let $\ve n$, $\ve p$ and $\ve q$ be an orthonormal coordinate system describing the orientation of a swimmer with $\ve n$ in its swimming direction and $\ve p$ in the direction perpendicular to $\ve n$ where the swimmer has most sensitive perception of flow disturbances. It could for instance be the direction of the antennae of a copepod, see Fig.~\ref{fig:mechanism}(a).
The symmetry breaking due to gravitational settling allows the swimmer to measure the $z$ component of $\ve p$ (assuming that gravity points in the negative $z$ direction) up to some resolution $\pzc$.
The strategy is to steer around its $\ve q$-axis with a positive angular velocity if $p_z>\pzc$ and with a negative angular velocity when $p_z<-\pzc$, both cases leading to the swimmer monotonously rotating towards larger $n_z$. Fig.~\ref{fig:mechanism}(b,c) illustrates this strategy for the case where the swimmer is unaffected by both fluid gradients and passive gyrotactic torque due to inhomogeneous mass distribution. Independent of the initial orientation, the swimmer can monotonously rotate towards larger $n_z$.
The strategy allows for both efficient migration upwards or downwards, not relying much on the physical characteristics of the swimmer in terms of shape, mass distribution and threshold value of the sensing.

In Section~\ref{sec:model}, we introduce the model for the microswimmers and our setup for reinforcement learning. In Section~\ref{sec:optimal}, we show results for the mechanism described above and compare it to swimmers using passive gyrotactics and refined active gyrotactic strategies relying on more flow signals. Conclusions and discussion of the results are presented in Section~\ref{sec:conclusions}.

\section{Model}
\label{sec:model}
We use a model similar to that in Ref.~\cite{Qiu2021}, but here we consider a three-dimensional flow.
We model the swimmer as a point particle with the shape of an elongated spheroid with aspect ratio $\lambda=a_{\parallel}/a_\perp$, where $a_\parallel$ is the axis length along the symmetry direction and $a_\perp$ is the transversal radius.
For typical microswimmers in the ocean, inertia of both the swimmer and the fluid can be neglected~\cite{Qiu2021}, leading to the following dynamics for the position $\ve x$, the symmetry direction $\ve n$, the transversal direction of the antennae $\ve p$, and the direction $\ve q$ perpendicular to the $\ve n$-$\ve p$ plane
\begin{subequations}
\label{eq:eqm}
\begin{align}
\label{eq:eqm_x}
\dot{\ve x}=\ve v
\,,\hspace{0.5cm}
\dot{\ve n}=\ve\omega\cross\ve n
\,,\hspace{0.5cm}
\dot{\ve p}=\ve\omega\cross\ve p
\,,\hspace{0.5cm}
\ve q=\ve n\cross\ve p
\,.
\end{align}
Here $\ve v$ and $\ve\omega$ are translational and angular velocity, given by~\cite{Durham2013,Qiu2021}
\begin{align}
\label{eq:eqm_v}
\ve v&=\ve u(\ve x,t)+\vvg+\vs\ve n\,,\\
\ve\omega&=\ve\Omega(\ve x,t)+\Lambda\ve n\cross(\ma S(\ve x,t)\ve n)+\frac{1}{2B}\ve n\cross\zzhat+\oomegas\,.
\label{eq:eqm_omega}
\end{align}
\end{subequations}
The flow velocity $\ve u$, half the flow vorticity $\ve\Omega=\ve\nabla\cross\ve u/2$ and the strain rate matrix $\ma S=(\ve\nabla\ve u+[\ve\nabla\ve u]^{\rm T})/2$ are evaluated at the instantaneous position of the swimmer. While $\ve u$ advects the swimmer's center of mass, the first two terms in $\ve\omega$ rotate it according to Jeffery's angular velocity for a spheroid with shape parameter $\Lambda=(\lambda^2-1)/(\lambda^2+1)$~\citep{Jef22}.

Assuming that the direction of gravity, $\gghat$, points in the negative $\zzhat$ direction, the swimmer settles with an orientation-dependent velocity~\cite{Kim:2005}
\begin{align}
\vvg=\left(\vgo-\vgp\right)n_z\ve n-\vgo\zzhat\,.
\label{eq:settling}
\end{align}
Here $\vgp$ and $\vgo$ are the velocities of a spheroidal particle settling in a quiescent flow with the symmetry axis parallel and perpendicular to gravity.
Assuming a bottom-heavy mass distribution, gravity also gives rise to a passive gyrotactic torque, turning the swimmer away from $\gghat$ at the time scale $B$~\cite{Kessler1985}, leading to the third term in $\ve\omega$.

Finally, the last term in $\ve v$ corresponds to swimming with constant speed $\vs$ in the instantaneous direction $\ve n$ and $\oomegas$  in $\ve\omega$ is an active angular velocity due to steering in response to measurements of the local environment.
We assume that $\omegasp=\oomegas\cdot\ve p$ and $\omegasq=\oomegas\cdot\ve q$ take either of the values $\{-\omegas,0,\omegas\}$, whereas $\omegasn=\oomegas\cdot\ve n$ remains zero.
Typical dimensional parameter values of copepods are shown in Table~\ref{tab:parameters}.

\subsection{Flow}
We use simulations of the dynamics (\ref{eq:eqm}) driven by a velocity field $\ve u$ obtained either by DNS or a statistical model for the flow velocity.
To make learning fast enough during training, we use frozen flow obtained by taking snapshots of the time-dependent flows in both the DNS and the statistical model. Later, when we evaluate the found strategies, we also consider the dynamics in time-dependent flows.
This approach of training in frozen flows only takes into account of spatial structure of the flow. We do not address the question of whether there exist even better strategies that exploit spatiotemporal structures in turbulence.

In our DNS we use the method of Ref.~\cite{Qiu2020a} to simulate incompressible homogeneous isotropic turbulence by numerical solution of Navier-Stokes equations
\begin{align}
\frac{\partial\ve u}{\partial t}+\ve u\cdot\ve\nabla\ve u=-\frac{\ve\nabla p^{(\rm f)}}{\rhof}+\nu\ve\nabla^2\ve u+\ve f\,,\hspace{0.5cm}\ve\nabla u=0
\label{eq:NS}
\end{align}
on a periodic domain.
Here $p^{(\rm f)}$, $\rhof$, and $\nu$ denote the pressure, density and kinematic viscosity of the fluid, respectively.
The turbulence is driven by an external force $\ve f$ with low wave number \cite{Machiels1997}. Eqs.~(\ref{eq:NS}) are solved by a pseudo-spectral method, and the $3/2$ rule is adopted to reduce the aliasing error on the nonlinear term. Time integration of both fluid and swimmer dynamics is performed using the explicit second-order Adam-Bashforth scheme~\cite{Rogallo1981}. Our simulations are carried out using $96^3$ grid points with Taylor scale Reynolds number ${\rm Re}=60$. The smallest resolved scale is about $1.78$ times smaller than the Kolmogorov scale, which lies within the accepted range to reach statistically reliable results~\cite{Pope2000}.

In the statistical model we use an incompressible Gaussian random velocity field on the form
\begin{equation}
    \ve{u}(\ve{x},t) = \frac{1}{\sqrt{6}}\nabla \times \ve{A}(\ve x,t)\,,
\label{eq:SM_u}
\end{equation}
where $\ve{A}$ is a three-dimensional vector potential with components being independent Gaussian random functions with zero mean and correlation function \cite{Gustavsson16review}
\begin{equation}
    \langle A_{i}(\ve x,\ve t)A_{j}(\ve x',\ve t')\rangle  = \delta_{ij}(\ell\urms)^2\exp\Big[-\frac{|\ve{x'}-\ve{x}|^2}{2\ell^2}-\frac{|t-t'|}{\tau}\Big].
\label{eq:SM_A}
\end{equation}
Here $\langle\cdot\rangle$ denotes a steady-state ensemble average, $\ell$ and $\tau$ are characteristic length and time scales, and $\urms=\langle \ve{u}^2 \rangle^{1/2}$.
The statistical model shows the best agreement with simulations in turbulent flows if large but finite values of $\ku=\urms\tau/\ell$ is considered. In this limit,
the flow evaluated at the positions of Lagrangian tracer particles decorrelates on the time scale $\ell/\urms$ due to displacement, rather than the time scale $\tau$ due to temporal variations of the flow. The relevant time scale in the statistical model in the limit of large $\ku$ is thus proportional to the Kolmogorov time $\tauK=\langle{\rm tr}(2\ma S^{\rm T}\ma S)\rangle^{-1/2}$, where $\ma S$ is the strain rate matrix. In the statistical model the Kolmogorov time evaluates to $\tauK=\ell/(\sqrt{5}\urms)$~\cite{Gustavsson16review}.

The root mean square velocity of turbulent fluctuations in the ocean, $\urms$, ranges from \SIrange{0.1}{100}{\milli\metre\per\s}~\cite{Yamazaki1996}.
The energy dissipation rate ranges from \SI{e-4}{\milli\metre\squared\per\s\cubed} in the deep sea, up to \SI{100}{\milli\metre\squared\per\s\cubed} in the upper ocean mixing layer~\cite{Yamazaki1996,Fuchs2016}.
For a kinematic viscosity of $\nu$ = \SI{1}{\milli\metre\squared\per\s}, the Kolmogorov time $\tauK$ takes values between \SI{0.1}{\s} and~\SI{100}{\s}.

\begin{table}	
	\begin{center}
	\begin{tabular}{llccc}\hline\hline
		&& range & used value & unit\\
				\hline
		Swimmer length  & $2a_\parallel$& 0.1-0.5& 0.2&\SI{}{\mm} \\
		Aspect ratio & $\lambda$ & 2.0-2.5& 2.0& \\
		Mass-density ratio & $\rhop/\rhof$& 1.005-1.019& 1.017& \\
		\multirow{2}*{Settling velocity} & $\vgp $&\multirow{2}*{0.1-0.8}& 0.15 & \SI{}{\mm\per\s}\\
		~& $\vgo $&~& 0.13 & \SI{}{\mm\per\s}\\
		Swimming velocity& $\vs$ & 0.33-3.76& 1.32 & \SI{}{\mm\per\s} \\
		Time scale of passive gyrotaxis & $B$& $\sim$ 10 & 5.0& \SI{}{\s}\\
		Swimming angular velocity & $\omegas$ & $<$ 20 & 1.14 & \SI{1}{\radian\per\s} \\
        Sensing threshold & $\Deltauc$ & $\ge$ 20 & 50 & \SI{}{\micro\metre\per\s}\\
				\hline\hline	
	\end{tabular}
	\caption{Typical values of model parameters estimated from juvenile copepods in the ocean and experiments, taken from Ref.~\cite{Qiu2021}.
The length is obtained for small copepods \citep{Titelman2001,Titelman2003} and the aspect ratio is estimated from their length and width~\citep{Carlotti2007}.
The mass-density ratio is obtained using mass density of copepods, $\rhop$, from Ref.~\citep{Knutsen2001} and a sea-water density of $\rhof =$ \SI{1.025}{\gram\per\cm\cubed} obtained for \SI{3.5}{\percent} salinity at \SI{20}{\degreeCelsius}, \citep{millero1980new}.
The settling velocity is obtained using the Stokes settling velocity for spheroids~\citep{Kim:2005} with our used values for $a_\parallel$, $\lambda$ and $\rhop/\rhof$.
The swimming velocity is taken from the experiments in Ref.~\citep{Titelman2003}.
The gyrotactic reorientation time of copepods is largely unknown. The value $B$ $\sim$ \SI{10}{\s} is taken from the critical vorticity in experiments on juvenile copepods~\citep{Fields1997}. In our simulations we use a slightly shorter reorientation time, $B$ = \SI{5}{\s}.
The maximal swimming angular velocity was estimated from the experiments in \citep{jiang2004}. In our model we use a smaller value that represents the slow horizontal and vertical steering motion described in Ref.~\citep{Kabata1971}.
The minimal sensing threshold $\Deltauc\sim \SI{20}{\micro\metre\per\s}$ is obtained from the measurements in Ref.~\cite{Yen1992}.
Simulation parameters are taken as the used values.
    }
	\label{tab:parameters}
	\end{center}
\end{table}

\subsection{Sensing and actions}
\label{sec:sensing}
Small microorganisms equipped with setae can use them to measure velocity differences between their body and the surrounding fluid~\cite{Strickler1973,Yen1992}.
Experiments show that copepods respond in different ways depending on the magnitudes of strain rate $\ma S$, angular velocity differences $\Delta\ve\Omega$, and velocity differences $\Delta\ve u$ of the flow~\cite{Kioerboe1999hydrodynamic}, indicating that they can distinguish between these three signals while not swimming.
This ability could be a consequence of that different flow structures give rise to different bending patterns of the copepod's setae~\cite{VisserBook,Kioerboe1999hydrodynamic}.
The disturbance to the surrounding flow by swimming complicates measurements~\cite{VisserBook}.
However, copepods are able to distinguish external hydrodynamic signals from its own generated flow when feeding~\cite{hwang2001can}.
It is therefore plausible that copepods are able to distinguish external hydromechanical signals from its own flow during steady swimming by recognizing spatial and temporal flow structures~\cite{yen1996advertisement} using an array of densely distributed setae along its body and antennae~\cite{Fields2014}.
Copepods often show a directional bias in their ability to sense hydromechanical signals~\cite{Fields2010}.
We assume a microorganism that is better at detecting signals in the symmetry direction $\ve n$ and the directions of its antennae $\pm\ve p$ than in the direction $\ve q$.
Primarily, we therefore consider navigation using the signals $\ma S$, $\Delta\ve\Omega$ and $\Delta\ve u$, projected on the $\ve n$ and $\ve p$ directions in the local frame of the swimmer. For completeness we also discuss strategies including the $\ve q$ direction because, in principle, a generic swimmer, which is able to use its setae to measure the flow at enough independent positions and orientations, has enough information to distinguish signals along both directions $\ve p$ and $\ve q$.

To reduce the number of signals, we use Eqs.~(\ref{eq:eqm}) to express velocity and angular velocity differences in the local coordinate system $\ve n$, $\ve p$ and~$\ve q$
\begin{subequations}
\label{eq:projections}
\begin{align}
\label{eq:vn}
\Delta u_n&=u_n-v_n=\vgp n_z-\vs\,,\\
\label{eq:vp}
\Delta u_p&=u_p-v_p=\vgo p_z\,,\\
\label{eq:vq}
\Delta u_q&=u_q-v_q=\vgo q_z\,,\\
\label{eq:omegan}
\Delta\Omega_n&=\Omega_n-\omega_n=-\omegasn\,,\\
\label{eq:omegap}
\Delta\Omega_p&=\Omega_p-\omega_p=\Lambda S_{nq}+\frac{1}{2B}q_z-\omegasp\,,\\
\label{eq:omegaq}
\Delta\Omega_q&=\Omega_q-\omega_q=-\Lambda S_{np}-\frac{1}{2B}p_z-\omegasq\,.
\end{align}
\end{subequations}
\noindent
Here subscripts $n$, $p$ or $q$ denote scalar product with the corresponding unit vector $\ve n$, $\ve p$ or $\ve q$.
First, according to Eq.~(\ref{eq:vn}), the magnitude $|\Delta u_n|\sim \vs$ is much higher than our used sensing threshold in Table~\ref{tab:parameters}.
To avoid introducing additional arbitrary threshold values, we skip this signal.
Second, Eqs.~(\ref{eq:vp}) and (\ref{eq:vq}) show that the swimmer is able to directly measure the $z$-component of the $\ve p$ and $\ve q$ directions and we adopt these components as signals.
Third, Eq.~(\ref{eq:omegan}) shows that measurement of $\Delta\Omega_n$ does not provide any information. Moreover, Eqs.~(\ref{eq:omegap}) and (\ref{eq:omegaq}) show that $\Delta\Omega_p$, is given by the signals $S_{nq}$ and $q_z$, and $\Delta\Omega_q$ is given by $S_{np}$ and $p_z$. Therefore $\Delta\ve\Omega$ does not contribute with any independent information and we neglect it as a signal.
Fourth, we only consider the strain components that directly affect the dynamics in Eqs.~(\ref{eq:projections}), $S_{np}$ and $S_{nq}$. Other strain components may give relevant information due to flow correlations, but this is most likely secondary to the direct contributions of $S_{np}$ and $S_{nq}$.
Finally, in our model we have omitted higher-order derivatives and time dependence of the hydromechanical signals.

The aim is to use reinforcement learning to search for optimal or approximately optimal strategies for vertical migration by suitable steering (actions) based on hydromechanical signals (states).
In our model we use different sets of states obtained from combinations of discretized signals $p_z$, $q_z$, $S_{np}$, and $S_{nq}$.
Each signal $\sigma$ is discretized into three states separated by a threshold level $\sigmac$
\begin{align}
\mbox{states for signal $\sigma$}=\left\{
\begin{array}{ll}
\sigma^{(+)} & \mbox{if }\sigma>\sigmac\cr
\sigma^{(0)} & \mbox{if }|\sigma|<\sigmac\cr
\sigma^{(-)} & \mbox{if }\sigma<-\sigmac
\end{array}
\right.\,.
\label{eq:state_discretization}
\end{align}
For the velocity differences $\Delta u_p$ and $\Delta u_q$, we use a threshold $\Deltauc$ = \SI{50}{\micro\metre\per\s}, giving the threshold $\pzc=\Deltauc/\vgo\approx 0.37$ for the $p_z$ and $q_z$ components with $\vgo$ from Table~\ref{tab:parameters}.
Estimating the corresponding gradients along the length of the swimmer, $\Deltauc/a_\parallel$~\cite{Kioerboe1999predator}, we obtain the threshold we use for angular velocity differences and strain rates, $\Sc$= \SI[per-mode=power]{0.5}{\per\s}.
We have chosen the threshold $\Deltauc$ to lie between the smallest velocity difference that copepods can physically sense, \SI{20}{\micro\metre\per\s}~\cite{Yen1992}, and the settling velocity $\sim$ \SI{150}{\micro\metre\per\s} in order to allow the copepod to sense its settling.
In experiments it is observed that different species of copepods make vigorous escape jumps in response to steady-flow strain rates of the order \SIrange[per-mode=power]{0.2}{20}{\per\s}~\cite{Kioerboe1999hydrodynamic,Kioerboe1999predator,Titelman2001,Buskey2002}.
Our threshold $\Sc$ is of this order of magnitude, indicating that it can be measured by copepods in nature.
We have confirmed that our results described in the next section are not sensitive to the exact values of the thresholds, as long as $\Deltauc$ is smaller than half the settling velocity.

We assume that the active angular velocity contribution, $\oomegas$ in Eq.~(\ref{eq:eqm_omega}), allows the swimmer to steer by rotating around the $\ve p$- and $\ve q$-axes with angular velocities $\omegasp$ and $\omegasq$ respectively.
We first consider a planar model, where the swimmer can measure signals in the $\ve n$-$\ve p$ plane: $p_z$ and $S_{np}$.
In this case we assume no steering around the $\ve p$-axis, $\omegasp=0$, and the angular velocity $\omegasq$ around the $\ve q$-axis takes three values, $\omegasq =\{-\omegas,0,+\omegas\}$.
This planar model is similar to the two-dimensional model considered in Ref.~\cite{Qiu2021}, but in that case $\ve q$ was fixed to the direction perpendicular to the flow plane, while $\ve q$ is free to rotate in the three-dimensional flow considered here.
We also consider a full three-dimensional model, where the swimmer in addition to the signals of the planar model has access to the signals $S_{nq}$ and $q_z$.
In addition to the three values of angular velocities around the $\ve q$-axis, the swimmer in the 3D model can also steer around the $\ve p$-axis with three levels $\omegasp = \{-\omegas,0,+\omegas\}$, giving in total $9$ different actions.

Copepods can acquire angular velocities up to \SI{20}{\radian\per\s} before a jump~\cite{jiang2004}. When cruising they steer with smaller angular velocities~\cite{Kabata1971}.
In our reinforcement learning we use $\omegas$ = \SI{1.14}{\radian\per\s}. We choose this value because it is of the order of the root-mean-square vorticity in our simulations during training, and it is much smaller than the hypothetical maximal angular velocity, $\vs/a_\parallel \approx$ \SI{13}{\radian\per\s}, that would have been obtained if the swimmer were able to convert the full swimming propulsion into angular rotation. Our value $\omegas$ = \SI{1.14}{\radian\per\s} gives a large length–specific turning radius, $\vs/(\omegas a_\parallel)\approx 12$, consistent with slow steering while cruising. We have confirmed that our results described in the next section are not sensitive to the exact value of $\omegas$, as long as it is not too small.

\subsection{Reinforcement Learning for vertical migration}

We adopt a one-step Q-learning algorithm~\cite{Watkins1992,Sutton1998,Mehlig2021} to train the swimmer to find efficient strategies for vertical migration.
The training is divided into a number of episodes.
In each episode, one precalculated frozen flow snapshot is randomly chosen. The swimmer starts with a random initial position and orientation and follows the dynamics~(\ref{eq:eqm}), first for an equilibration time $\Teq$, and then for a predefined physical time of training, $\Ttrain$.
The initial equilibration is introduced to search for strategies that are efficient in the statistical steady state, not relying on the arbitrary uniform initial condition.
In the DNS at regularly distributed time steps $i\Tmeas$ with $i=0,1,\dots,\lfloor(\Teq+\Ttrain)/\Tmeas\rfloor$, the swimmer measures its current discrete state $s_i$, obtained by measurement of the signals described above, and is given a reward $r_i$.
In the statistical model a reward is instead given at any time step where the state changes.
Because the goal is vertical migration upwards, we use the velocity in the $z$ direction since the last state measurement as reward in the DNS, $r_i = (z_{i+1} - z_{i})/\Tmeas$, and displacement, $r_i = z_{i+1} - z_i$, in the statistical model.
After the reward is given, the swimmer sets $\oomegas$ to action $a_i$.
During the equilibration, $T<\Teq$, the action is chosen according to a greedy policy, $a_i={\rm arg\,max}_a Q(s_i,a)$ and the Q-table is kept constant. After the equilibration the action is chosen according to an $\varepsilon$-greedy policy
\begin{align}
\label{eq:policy}
a_i=\left\{
\begin{array}{ll}
\mbox{random action} & \mbox{with probability }\varepsilon\cr
{\rm arg\,max}_a Q(s_i,a) & \mbox{otherwise}
\end{array}
\right.\,,\hspace{0.5cm}
\mbox{where }
\varepsilon=\varepsilon_0\,{\rm max}\left(0,1-\frac{E}{\Eeps}\right)\,.
\end{align}
Here $\varepsilon$ is an exploration rate that starts at $\varepsilon_0$ and decays linearly with the training episode number $E$ until it reaches zero at episode $\Eeps$.
It allows the swimmer to explore different actions, preventing the learning to get stuck at local optima.
Normally, non-exploratory actions are chosen as the action giving the maximal value of the $Q$-table, $Q(s,a)$, for the current state.
Starting with a $Q$-table of uniform values, it is updated each time the agent is given a reward using the rule
\begin{equation}\label{qlearning}
Q(s_i,a_i) \leftarrow Q(s_i,a_i) + \alpha[r_i + \gamma \max_{a}Q(s_{i+1},a) - Q(s_i,a_i)]\,,\hspace{0.5cm}
\mbox{where }
\alpha=\alpha_0\frac{\Ealpha}{\Ealpha+E}\,.
\end{equation}
Here $\alpha$ is a learning rate, starting at $\alpha_0$ and decaying on the episode scale $\Ealpha$.
The discount factor $\gamma$, with $0\le\gamma<1$, sets an optimization time scale over $(1-\gamma)^{-1}$ state changes. It is introduced to prevent divergence of the values in the $Q$ table in the long run. We adopt a far-sighted optimization by setting $\gamma$ close to $1$.
At the end of each episode, the $Q$ table is kept for the next episode, continuing to be updated.
In the ideal case of infinitely many exploratory moves and if the system is Markovian, the entries of the $Q$-table approach the optimal values for the expectation values of the future discounted reward, $Q(s_i,a)=\langle\sum_{j=0}^\infty \gamma^j r_{i+j}\rangle$ for taking action $a$ in state $s_i$ following the policy~(\ref{eq:policy}). In most realistic situations, the process instead converges to an approximately optimal strategy.
Information about the parameters used in our different training cases are given in Appendix~\ref{appendix:learning_parameters}.

\section{Optimal swimming strategies}
\label{sec:optimal}

\subsection{Results from reinforcement learning}
\begin{table}[]
	\begin{tabular}{b{1.8cm}b{1.8cm}b{2.5cm}p{0.8cm}p{1.4cm}p{0.8cm}p{1.4cm}l}
	& & & \multicolumn{4}{c}{Performance $\langle v_z\rangle/\langle v_z\rangle_0$\vspace*{0.1cm}}\\
	Model & Actions & Signals & \multicolumn{2}{l}{Frozen DNS} & \multicolumn{2}{l}{Frozen statistical model} \\
	  \hline
    Naive & $-$ & $-$ &  0.30 & & 0.22 & \\
	  \hline
    Planar & $\omegasq$ & $p_z$ & 0.58 & (0.58) & 0.73 & (0.73) & (Eq. (\ref{eq:optimal}))\\
	\hspace{0.45cm}" & \hspace{0.1cm}" & $S_{np}$ & 0.31 & & 0.22 & &\\
	\hspace{0.45cm}" & \hspace{0.1cm}"& $p_z$, $S_{np}$ & 0.58 & & 0.73 & & \\
	  \hline
    Full 3D & $\omegasp$, $\omegasq$ & $p_z$ & 0.58 & & 0.73 & & \\
	\hspace{0.45cm}"& \hspace{0.45cm}" & $q_z$ & 0.58 & & 0.73 & & \\
	\hspace{0.45cm}"& \hspace{0.45cm}"& $p_z$, $q_z$ & 0.64 & (0.66) & 0.94 & (0.94) & (Eq. (\ref{eq:optimal_two_signals})) \\
	\hspace{0.45cm}"& \hspace{0.45cm}"& $S_{np}$ & 0.32 & & 0.22 & & \\
	\hspace{0.45cm}"& \hspace{0.45cm}"& $S_{nq}$ & 0.34 & & 0.22 & & \\
	\hspace{0.45cm}"& \hspace{0.45cm}"& $S_{np}$, $S_{nq}$ & 0.31 & & 0.20 & & \\
	\hspace{0.45cm}"& \hspace{0.45cm}"& $p_z$, $S_{np}$ & 0.57 & & 0.73 & & \\
	\hspace{0.45cm}"& \hspace{0.45cm}"& $p_z$, $S_{nq}$ & 0.58 & & 0.73 & & \\
	\hspace{0.45cm}"& \hspace{0.45cm}"& $q_z$, $S_{np}$ & 0.57 & & 0.73 & & \\
	\hspace{0.45cm}"& \hspace{0.45cm}"& $q_z$, $S_{nq}$ & 0.58 & & 0.73 & & \\
	\hspace{0.45cm}"& \hspace{0.45cm}"& $p_z$, $q_z$, $S_{np}$, $S_{nq}$ & 0.64 & (0.68) & 0.94 & (0.94)  & (Eq. (\ref{eq:optimal_four_signals})) 	\end{tabular}
\caption{\label{tab:DNS_signal_comb}Average vertical velocity $\langle v_z\rangle$ normalized by the value in quiescent flows $\langle v_z\rangle_0$ [Eq.~(\ref{eq:vz0})] for different sets of discretized actions and states, see Section~\ref{sec:sensing}, with parameters from Table~\ref{tab:parameters}. Numbers show results obtained by evaluation of the best strategy found for each case using reinforcement learning in frozen DNS or statistical model flows. The numbers in parentheses show results for the strategies in Eqs.~(\ref{eq:optimal}), (\ref{eq:optimal_two_signals}) and (\ref{eq:optimal_four_signals}). Errors of the numerical values are of the order $\pm 0.01$.
}
\end{table}
We evaluate the performance of the strategies found in our reinforcement learning by running the same setup as in the training, but with a greedy action ($\varepsilon=0$) and we do not update the $Q$-table ($\alpha=0$). We average the vertical velocity over multiple episodes, each of time $\Ttrain$ sampled after the initial equilibration time $\Teq$. Both $\Teq$ and $\Ttrain$ are chosen much larger than the maximal time scale of the dynamics, so that steady state statistics is obtained.
The results of the best strategies found are summarized in Table~\ref{tab:DNS_signal_comb}.
The vertical velocity is normalized using the average
\begin{align}
\langle v_z\rangle_0=\vs - \vgp\,,
\label{eq:vz0}
\end{align}
obtained in the long-time limit if the ambient flow velocity $\ve u$ is put to zero in Eq.~(\ref{eq:eqm_v}).
Results for the naive model, obtained for a swimmer with zero active angular velocity, $\oomegas=0$, are given as a reference.

\begin{figure}
\includegraphics[width=15.5cm]{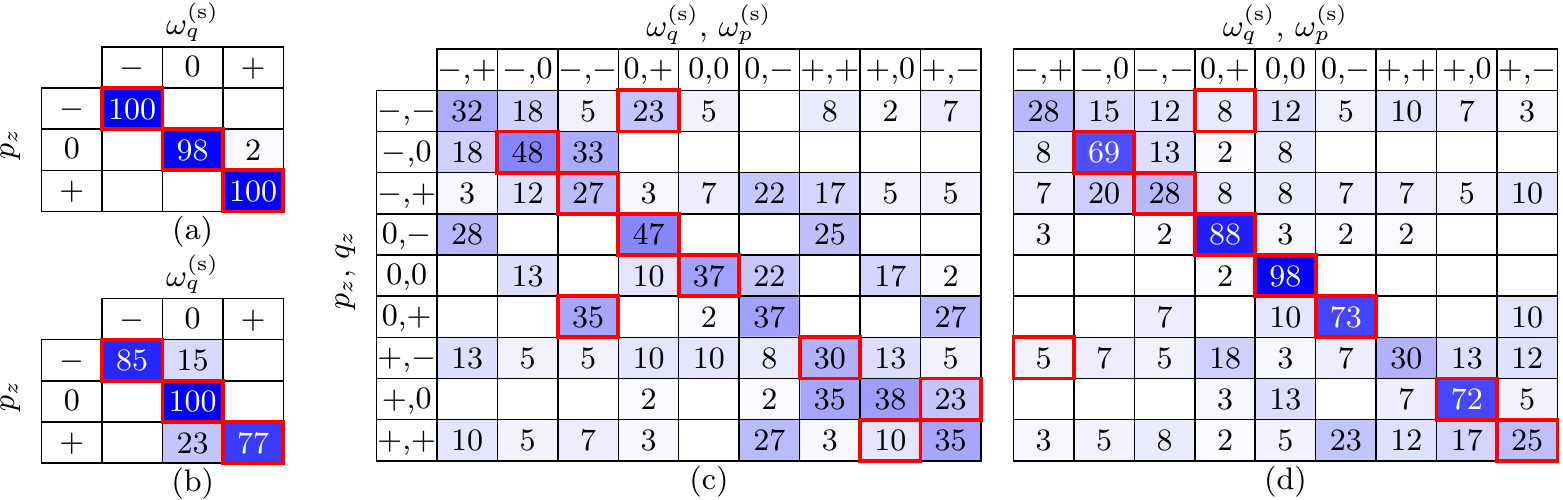}
    \caption{Summary of strategies obtained by reinforcement learning for the cases of the planar model with single signal $p_z$ and full 3D model with joint signal $p_z$ and $q_z$. Results from the frozen DNS [(a) and (c)] and the statistical model [(b) and (d)]. Rows represent different states, the signs $-$, $0$, and $+$ represent the states $\sigma^{(-)}$, $\sigma^{(0)}$, and $\sigma^{(+)}$ in the discretized signal in Eq.~(\ref{eq:state_discretization}). Columns represent different actions, where the signs represent the values $-\omegas$, $0$, and $+\omegas$ respectively. The action selected in a given state for the best strategy found in our reinforcement learning is highlighted in red. In each panel the resulting strategies from 60 completed training sessions are summarized. Each numerical value gives the non-zero percentage of strategies choosing a certain action in a given state. The background is color coded according to this percentage from white (0\%) to blue (100\%).
    }
	\label{fig:Policy_planar_DNS}
\end{figure}

In the planar model the swimmer can steer around the $\ve q$-axis, $\oomegas=\omegasq\ve q$, and it can sense combinations of two signals: the vertical component $p_z$ of the $\ve p$-axis and the strain component $S_{np}$.
The results in Table~\ref{tab:DNS_signal_comb} show that the signal $p_z$ is more important for upward vertical migration than $S_{np}$ in both the DNS and the statistical model.
The optimal strategy when using only $p_z$ as the signal is the same in the two flows, and it is highlighted using red frames in Fig.~\ref{fig:Policy_planar_DNS}(a,b).
Fig.~\ref{fig:mechanism} shows the mechanism explaining why this strategy is successful.
The optimal strategy when using only $S_{np}$ as the signal is on the same level as the naive strategy. The optimal strategy using both $p_z$ and $S_{np}$ performs approximately on the same level as the simpler strategy using only $p_z$ as the signal.

In the full 3D model, the swimmer can rotate around both the $\ve p$- and $\ve q$-axes, and all four relevant flow signals are considered.
Using a single signal $p_z$, the best strategy found for both the DNS and statistical model is the same as in the planar case, i.e. rotate around the $\ve q$-axis according to Fig.~\ref{fig:Policy_planar_DNS}(a), with no rotation around the $\ve p$-axis. Using $q_z$ as the signal, the equivalent strategy is found by rotating around $\ve p$ with no rotation around $\ve q$.
For two combined signals, the best strategies found in both DNS and statistical model are based on $p_z$ and $q_z$. They are highlighted using red frames in Fig.~\ref{fig:Policy_planar_DNS}(c,d).

The strategies based only on the strain signals perform on the level of the naive swimmer, with approximately half the vertical velocity compared to the strategies based on $p_z$ and $q_z$. Hybrid strategies between one strain component and either of $p_z$ or $q_z$ perform on the same level as the individual $p_z$ or $q_z$ signal.
Finally, in the case of four signals, $p_z$, $q_z$, $S_{np}$, and $S_{nq}$, the best strategies found are of the same order as the strategies based on $p_z$ and $q_z$ only.

Being a stochastic process, different training sessions often converge to different approximately optimal strategies. Therefore, we have run several training sessions for each case considered, and the results displayed in Table~\ref{tab:DNS_signal_comb} are based on the strategy with the best performance during training.
The cell colors in Fig.~\ref{fig:Policy_planar_DNS} summarize the resulting strategies from 60 training sessions for each case (a--d).
For a given state, an action with a white or light-blue cell are rarely chosen, while actions with blue color are selected in most of the resulting strategies.
We find that for a single signal in the planar model, most training sessions converge to the optimal strategy highlighted in red in Fig.~\ref{fig:Policy_planar_DNS} (a,b).
For two signals in the full 3D model, results are more scattered.
The reason is that many strategies have similar reward levels and it is therefore hard to find the global optimal strategy.
In the DNS the mean value of $\langle v_z\rangle/\langle v_z\rangle_0$ obtained in the 60 training sessions is $0.60$ and the median is $0.61$, both being close to the value $0.64$ for the best strategy found. The same holds for the statistical model where the mean value is $0.85$ and the median is 0.94, equal to the value for the best strategy found.
We conclude that although it is hard to find the global optimum, many quasi-optimal strategies have approximately the same performance.

The strategies discussed above are obtained in frozen flows. The flow is changed each episode to reduce bias towards specific frozen flows.
However, the flow statistics of the dynamics in a frozen flow is slightly different from the statistics in time fluctuating flows, or for swimmers with other parameter values.
In general, which strategy is optimal for a given task depends on both the model parameters and learning parameters, such as the choice of actions and states. There is no guarantee that a strategy found for one setup is also the best strategy for a different setup. For example, the optimal strategy found using a frozen flow is not necessarily the same as the optimal strategy in a time-dependent flow, and a flow in nature could have yet another optimal strategy.
We therefore try to identify generally valid strategies that are robust upon changing the details.
Using the trend that the best strategies in Fig.~\ref{fig:Policy_planar_DNS} tend to chose actions along the diagonals, we formulate strategies below in the planar model and full 3D model that have good performance for a large number of parameters and flows, including time-dependent flows.

\subsection{Planar model with a single signal}
Our reinforcement learning shows that for upward vertical navigation in the planar model using a single observable and the action to steer around the $\ve q$-axis, the most important signal is $p_z$, being proportional to the cosine of the angle between the transversal direction $\ve p$ and the direction of gravity $\gghat=-\zzhat$.
The resulting optimal strategy, shown in Fig.~\ref{fig:Policy_planar_DNS}(a,b), can be written mathematically as $\oomegas=\omegasp\ve p+\omegasq\ve q$ with
\begin{align}
\omegasp=0
\hspace{0.5cm}
\mbox{and}
\hspace{0.5cm}
\omegasq=\omegas
\left\{
\begin{array}{ll}
1&\mbox{ if }p_z>\pzc\cr
-1&\mbox{ if }p_z<-\pzc\cr
0 & \mbox{ otherwise}
\end{array}
\right.\,.
\label{eq:optimal}
\end{align}
This strategy imposes an active gyrotactic stability on the orientational dynamics.
To understand the mechanism of Eq.~(\ref{eq:optimal}), we consider a simplified dynamics where passive gyrotaxis and flow gradients are neglected. In this limit, the angular velocity of the swimmer, Eq.~(\ref{eq:eqm_omega}), is solely determined by the active swimming contribution, $\ve\omega=\oomegas$, and the orientational dynamics of Eq.~(\ref{eq:eqm_x}) becomes
\begin{align}
\dot{\ve n}=\omegasq\ve p
\,,\hspace{0.3cm}
\dot{\ve p}=-\omegasq\ve n
\hspace{0.3cm}\mbox{and}\hspace{0.3cm}
\dot{\ve q}=0\,.
\label{eq:mechanism}
\end{align}
Assuming $\omegas>0$, the signature of $\omegasq$ only depends on $p_z$ according to Eq.~(\ref{eq:optimal}).
If the initial value $p_{z,0}$ is smaller than the sensing threshold, $|p_{z,0}|\le\pzc$, the orientation remains unchanged. If instead $|p_{z,0}|>\pzc$, the active rotational swimming turns the swimmer around its $\ve q$-axis until $p_z$ reaches $\pzc$ for the case of positive $p_{z,0}$, or until it reaches $-\pzc$ for the case of negative $p_{z,0}$. The direction of the rotation following from Eq.~(\ref{eq:optimal}) is such that the $z$ component of $\ve n$ in Eq.~(\ref{eq:mechanism}) is rotated towards larger values, $\dot n_z=\omegas|p_z|\Theta(|p_z|-\pzc)\ge 0$, where $\Theta$ is the Heaviside step function. Fig.~\ref{fig:mechanism}(b,c) illustrate this mechanism for the case $\pzc=0.37$.

To evaluate the average alignment $\langle n_z\rangle$ for the simplified dynamics (\ref{eq:mechanism}), we use the normalisation $n_z^2+p_z^2+q_z^2=1$.
If $|p_{z,0}|\le \pzc$, there is no rotation and the final value $n_{z,{\rm f}}$ is equal to the initial value $n_{z,0}$.
If $|p_{z,0}|> \pzc$, then $q_z$ remains constant, while $p_z$ is rotated until the threshold $\pm\pzc$.
Using the normalisation, the final value of $n_z$ must take the form $n_{z,{\rm f}}=\pm\sqrt{1-[\pzc]^2-q_{z,0}^2}$.
The sign choice in Eq.~(\ref{eq:optimal}) breaks symmetry, always giving the positive solution of $n_{z,{\rm f}}$, while the negative solution would be obtained by rotating in the opposite direction of Eq.~(\ref{eq:optimal}).
Averaging the positive solution $n_{z,{\rm f}}$ over initially uniformly distributed orientations, assuming $0\le\pzc\le 1$, gives (see Appendix \ref{appendix:alignment} for details)
\begin{align}
\label{eq:nz_planar}
\langle n_z\rangle=\frac{4}{3\pi}\frac{(1-[\pzc]^2)^{3/2}}{\pzc}{}_3F_2\left[\frac{1}{2},1,2;\frac{3}{2}, \frac{5}{2};1 - \frac{1}{[\pzc]^2}\right]\,.
\end{align}
Here ${}_3F_2$ is the generalized hypergeometric function.
This solution scales as $\langle n_z\rangle\sim\tfrac{\pi}{4}-\tfrac{2}{\pi}\pzc-\tfrac{\pi}{4}[\pzc]^2$ for small $\pzc$.
The limiting case of very high resolution of the signal, $\pzc\to 0$, results in large albeit not perfect alignment, $\langle n_z\rangle\to\tfrac{\pi}{4}\approx 0.8$.
As $\pzc$ approaches unity, the solution scales as $\langle n_z\rangle\sim \tfrac{8\sqrt{2}}{3\pi}(1-\pzc)^{3/2}$. For $\pzc>1$, the swimmer cannot resolve $p_z$ and active alignment fails.
We conclude that Eq.~(\ref{eq:mechanism}) is a simple mechanism for active gyrotactic stability, leading to partial alignment against gravity in quiescent flows even without inhomogeneous mass distribution.

\begin{figure}
\includegraphics[width=17.5cm]{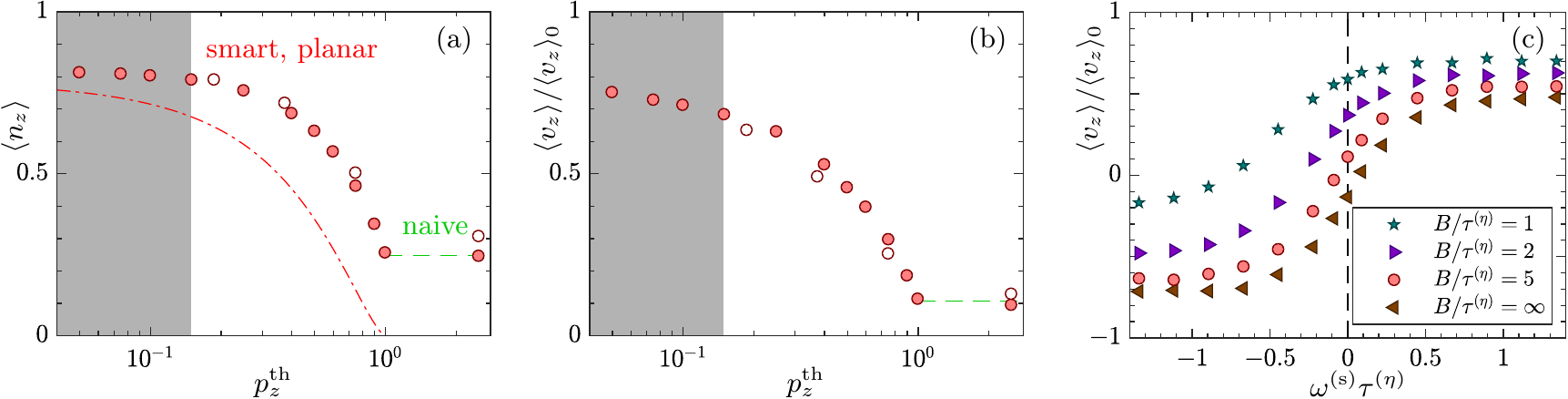}
\caption{
\label{fig:velocity}
Numerical simulation results in the steady state for average alignment (a) and velocity (b,c) against the threshold of the flow signal, $\pzc=\Deltauc/\vgo$ (a,b), and the dimensionless angular swimming velocity $\omegas\tauK$ (c).
Velocities are normalized using $\langle v_z\rangle_0$ in Eq.~(\ref{eq:vz0}).
Symbols show results from simulations of Eqs.~(\ref{eq:eqm}) with smart planar steering [$\oomegas=\omegasq\ve q$ using Eq.~(\ref{eq:optimal})] in the statistical model with $\ku=10$ (colored symbols) and DNS [hollow symbols in panels (a,b)].
Numerical results for the naive strategy ($\oomegas=0$) in the statistical model are shown as horizontal dashed green lines.
The shaded region corresponds to threshold levels below the minimal resolution limit $\Delta u$ = \SI{20}{\micro\metre\per\s}.
The dash-dotted line in panel (a) shows the theory for the simplified dynamics, Eq. (\ref{eq:nz_planar}).
Flow parameters $\urms$ = \SI{6.7}{\mm\per\s} and $\tauK$ = \SI{1}{\s}. All swimmer parameters according to the used values in Table~\ref{tab:parameters}, except for parameter values stated in the figure.}
\end{figure}

Fig.~\ref{fig:velocity} shows simulation results for the full dynamics~(\ref{eq:eqm}) using the strategy (\ref{eq:optimal}) with parameters from Table~\ref{tab:parameters}. The data illustrates the sensitivity of the strategy to the choice of the dimensionless parameters $\pzc$, $\omegas\tauK$ and $B/\tauK$.
Fig.~\ref{fig:velocity}(a) shows the dependence of the average vertical alignment $\langle n_z\rangle$ on the signal threshold $\pzc$.
As expected, for $\pzc>1$ the alignment is equal to the naive case with no steering, $\omegas=0$. In this case there is a weak alignment due to the passive gyrotactic angular velocity in Eq.~(\ref{eq:eqm_omega}).
When $\pzc$ is reduced below unity, there is a sharp increase in $\langle n_z\rangle$ until $\pzc\approx 0.5$. For smaller values of $\pzc$, $\langle n_z\rangle$ is approximately constant around $0.8$. We conclude that the results are not sensitive to the value of $\pzc$, as long as it is smaller than $0.5$.
Our choice of $\pzc=0.37$ lies in the upper part of this range.
The dash-dotted curve in Fig.~\ref{fig:velocity}(a) shows the analytical evaluation for a quiescent flow, Eq.~(\ref{eq:nz_planar}).
It has the same trend as the numerical data, but predicts a somewhat smaller alignment. This is expected because the passive gyrotactic contribution in the numerical simulations adds to the alignment and the turbulent velocity gradients do not affect the alignment too much for the flow with $\urms$ = \SI{6.7}{\mm\per\s} and $\tauK$ = \SI{1}{\s} in Fig.~\ref{fig:velocity}.
Fig.~\ref{fig:velocity}(b) shows that the same conclusion holds for the average vertical velocity $\langle v_z\rangle$.
Starting at the naive result for $\pzc>1$, $\langle v_z\rangle$ increases quickly until $\pzc\sim 0.5$, where a plateau is reached with velocity around $0.7\langle v_z\rangle_0$, with $\langle v_z\rangle_0$ given in Eq.~(\ref{eq:vz0}).
Hollow markers in Fig.~\ref{fig:velocity}(a,b) show results from our DNS. Since they qualitatively agree with the statistical model, we limit the analysis in what follows to the statistical model which is quicker to evaluate.

Fig.~\ref{fig:velocity}(c) shows the vertical velocity component for statistical model simulations against $\omegas\tauK$ for different values of $B/\tauK$.
The data reaches plateaus for $|\omegas|\tauK>0.8$, meaning that the results are not sensitive to the choice of the dimensionless steering angular velocity if it is large enough.
For  $\tauK$ = \SI{1}{\s}, the choice $\omegas\tauK=1.14$ in Table~\ref{tab:parameters} lies well within the plateaus.
When $B/\tauK\to\infty$ (brown,$\triangleleft$), the passive gyrotactic torque vanishes, and by solely using the active gyrotactic mechanism (\ref{eq:optimal}), the swimmer reaches an upward velocity of about $0.5\langle v_z\rangle_0$.
For negative values of $\omegas$, the active mechanism works in the opposite direction, giving alignment with gravity and a means to migrate downwards with a slightly larger speed due to the contribution from settling.
Adding passive gyrotaxis with reorientation time $B/\tauK=5$ from Table~\ref{tab:parameters} gives an additive contribution of around $0.1\langle v_z\rangle_0$ for the full range of $\omegas\tauK$ (red,$\circ$).
To modify the vertical velocity significantly, the reorientation time must be of the order of $\tauK$ (green,$\star$) for the flow considered here.
Results for passive gyrotaxis is obtained along the line $\omegas=0$ in Fig.~\ref{fig:velocity}(c).
When $B/\tauK>2$, the active gyrotactic stability is dominant, while for $B/\tauK<2$ passive gyrotaxis is efficient for upward migration and the active mechanism only give a minor contribution.

In conclusion, our numerical simulations show that the good performance of the found strategy is not sensitive to the precise values of the parameters or the statistics of the flow used in training.
For the parameters considered in Fig.~\ref{fig:velocity}, vertical alignment due to passive and active gyrotactic reorientation result in high vertical velocities.
Both the average flow velocity $\langle u_z\rangle$ and the settling velocity $\vvg$ in Eq.~(\ref{eq:eqm_v}) are at most of order $0.1\langle v_z\rangle_0$, which is small in comparison.
We discuss the relative contributions of the terms in Eq.~(\ref{eq:eqm_v}) in more detail below.
In what follows we consider the parameters in Table~\ref{tab:parameters}. For these, the active gyrotactic mechanism dominates over the passive one.

\subsection{Full 3D model}
In our 3D model the swimmer can steer around both $\ve q$ and $\ve p$, and respond to signals in both directions with the threshold values $\Deltauc$ for velocity differences and $\Sc$ for fluid gradients.
Our reinforcement learning shows that when only one signal is used, the dominant signal is still $p_z$ and the strategy (\ref{eq:optimal}) remains optimal. There also exists an equivalent strategy where $q_z$ is the signal and the agent steers around $\ve p$ instead of $\ve q$.
It gives rise to the same mechanism for aligning $\ve n$ against gravity as in Fig.~\ref{fig:mechanism}, but from rotations around the $\ve p$-axis with $q_z$ as signal.

For swimmers allowed to use two signals when navigating, the best strategies obtained from reinforcement learning use the combined signal of $p_z$ and $q_z$, see Table~\ref{tab:DNS_signal_comb}.
They are highlighted in Fig.~\ref{fig:Policy_planar_DNS}(c,d). A common trend in the DNS and the statistical model is that the best strategies frequently take actions along the diagonal.
It is therefore of interest to compare to the strategy of choosing actions along the diagonal in Fig.~\ref{fig:Policy_planar_DNS}(c,d).
This diagonal strategy is simply a superposition of the two single-signal strategies described above, which can be expressed mathematically as $\oomegas=\omegasp\ve p+\omegasq\ve q$ with
\begin{align}
\omegasp=-\omegas
\left\{
\begin{array}{ll}
1&\mbox{ if }q_z>\pzc\cr
-1&\mbox{ if }q_z<-\pzc\cr
0 & \mbox{ otherwise}
\end{array}
\right.
\hspace{0.5cm}
\mbox{and}
\hspace{0.5cm}
\omegasq=\omegas
\left\{
\begin{array}{ll}
1&\mbox{ if }p_z>\pzc\cr
-1&\mbox{ if }p_z<-\pzc\cr
0 & \mbox{ otherwise}
\end{array}
\right.\,.
\label{eq:optimal_two_signals}
\end{align}
Here $\omegasq$ is identical to that in Eq.~(\ref{eq:optimal}) and $\omegasp$ takes the same expression with $p_z$ replaced by $q_z$ and multiplied by a minus sign as a consequence of the relative handedness between the vectors $\ve n$, $\ve p$ and $\ve q$.
Corresponding to our simulations, we have chosen the same threshold level $\pzc$ for both $p_z$ and $q_z$, and that the swimmer turns around the axes $\ve p$ and $\ve q$ at the same angular rate $\omegas$.
We find that the performance of strategy (\ref{eq:optimal_two_signals}) is slightly better than the optimal policy obtained by reinforcement learning, see Table~\ref{tab:DNS_signal_comb}.
This is a consequence of the reinforcement learning getting stuck in local optima.
Although we cannot be certain that strategy~(\ref{eq:optimal_two_signals}) is the global optimal strategy for our flows, we expect it to be close to optimal and we analyze its dynamics below.

Strategy (\ref{eq:optimal_two_signals}) gives rise to terms that adds to the contributions of the two passive gyrotactic terms in $\omega_p=\Omega_p-\Delta\Omega_p$ and $\omega_q=\Omega_q-\Delta\Omega_q$ in Eqs.~(\ref{eq:omegap}) and (\ref{eq:omegaq}), and therefore strengthen the gyrotactic contribution.
Using the simplified dynamics (no flow gradients or passive gyrotaxis) used to derive Eq.~(\ref{eq:nz_planar}), the alignment in the present case becomes $\langle n_z\rangle\sim 1-(1+\tfrac{2}{\pi})[\pzc]^2$ for small $\pzc$ and $\langle n_z\rangle\sim \tfrac{16\sqrt{2}}{3\pi}(1-\pzc)^{3/2}$ for $\pzc$ close to unity, see Appendix~\ref{appendix:alignment} for details.
In contrast to the case of a single signal, the present strategy allows for perfect alignment in quiescent flows if the threshold level is small enough. The alignment has a plateau for small $\pzc$ and decreases sharply to zero as $\pzc$ approaches unity, similar to Eq.~(\ref{eq:nz_planar}) shown as the dash-dotted line in Fig.~\ref{fig:velocity}(a).

\begin{figure}
\includegraphics[width=17.5cm]{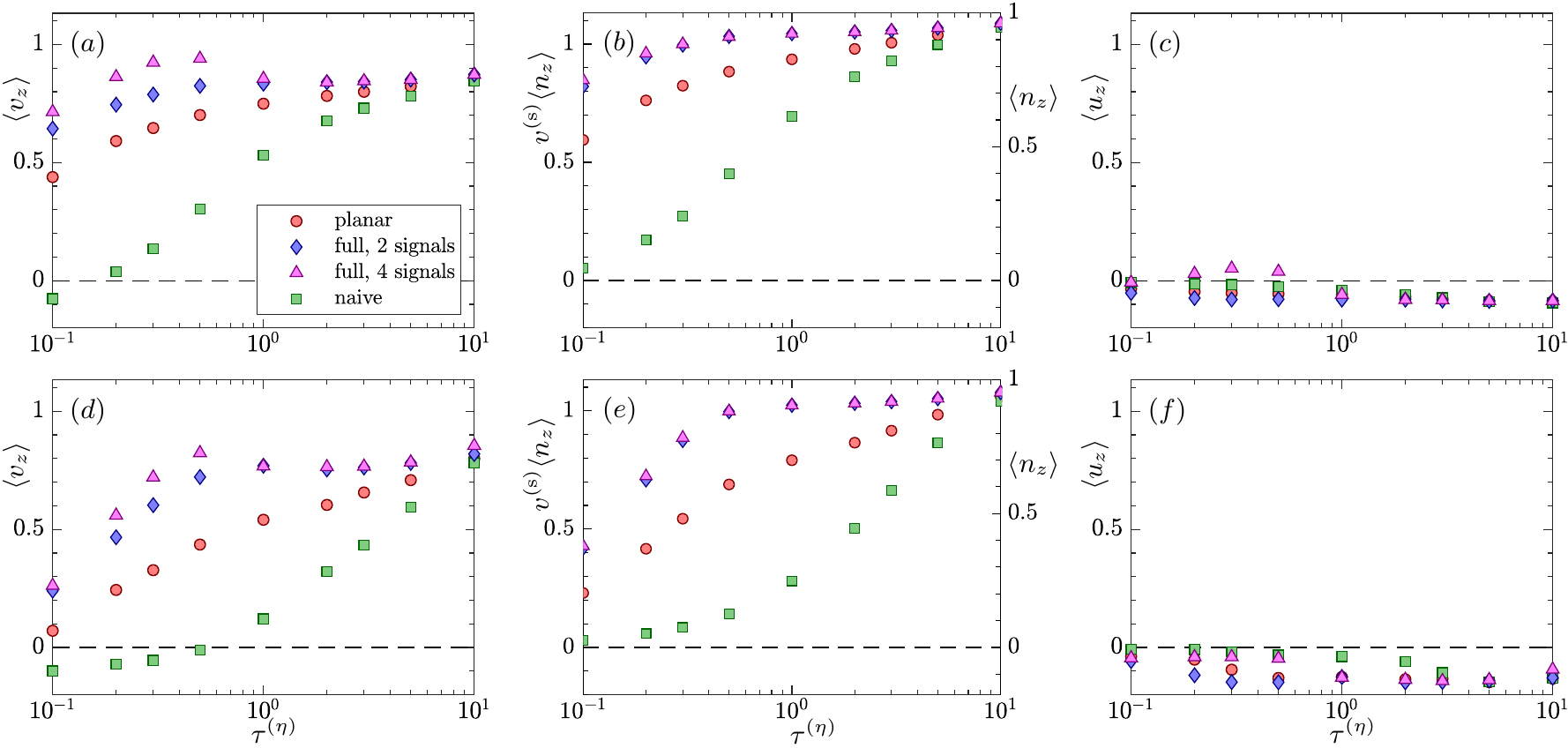}
\caption{
\label{fig:velocity_full}
Left column: Steady-state results from statistical model simulations for the average vertical velocity $\langle v_z\rangle$ against the Kolmogorov time $\tauK$ (units of seconds)
for a swimmer following either the planar strategy~(\ref{eq:optimal}) [red,$\circ$], the full 3D strategy using two signals (\ref{eq:optimal_two_signals}) [blue, $\Diamond$], the full 3D strategy using four signals (\ref{eq:optimal_four_signals}) [magenta, $\vartriangle$], or the naive passively gyrotactic strategy (green, $\Box$).
Parameters of the swimmer are given in Table~\ref{tab:parameters}. The flow velocity is $\urms$ = \SI{1}{\milli\metre\per\s} (top row) and $\urms$ = \SI{10}{\milli\metre\per\s} (bottom row) with a constant Kubo number, $\ku=10$.
Middle column: Same for the vertical velocity due to swimming, $\vs\langle n_z\rangle$ (the scale on the right axis shows the alignment $\langle n_z\rangle$).
Right column: Same for the vertical component of the flow velocity along swimmer trajectories, $\langle u_z\rangle$.
The displayed velocities are normalized by $\langle v_z\rangle_0$ in Eq.~(\ref{eq:vz0}).
}
\end{figure}

For the case of all four signals, $p_z$, $q_z$, $S_{np}$, and $S_{nq}$, it is harder to read off a simple strategy from our reinforcement learning results.
Since the best strategies found do not surpass the results of strategy~(\ref{eq:optimal_two_signals}), see Table~\ref{tab:DNS_signal_comb}, we do not expect that the reinforcement learning has converged to the global optimal strategy.
This is no surprise: with 9 actions and $81$ states considered here, the number of possible strategies is enormous, of the order $10^{77}$.
We therefore proceed in a different way to find an efficient and interpretable strategy.
In the strategy (\ref{eq:optimal_two_signals}), the signals $p_z$ and $q_z$ are used to strengthen the passive gyrotaxis.
In \cite{Qiu2021}, $S_{np}$ was used to increase the shape factor $\Lambda$ to a larger effective value, making the swimmers preferentially sample up-welling regions, which facilitates upward migration.
This was achieved by choosing $\omegasq=\omegas{\rm sign}(S_{np})$ when $|S_{np}|>\Sc$ and 0 otherwise, i.e. adding a contribution to the angular velocity with the same sign as $S_{np}$, effectively increasing $\Lambda$ by $\omegas$ when $|S_{np}|$ is above the threshold value.
We choose a strategy that is a superposition of the two effects of strengthening the passive gyrotactic angular velocity and increasing the effective shape factor.
Testing different combinations of putting priority to either $\{p_z,q_z\}$, $\{S_{np},S_{nq}\}$, or a combination thereof, shows that prioritizing $\{p_z,q_z\}$ works best, i.e. use the strategy (\ref{eq:optimal_two_signals}) when the $p_z$- or $q_z$-signal is above its threshold and refine this strategy using $\{S_{np},S_{nq}\}$ when the signal is below the threshold:
\begin{align}
\omegasp=-\omegas
\left\{
\begin{array}{ll}
1&\mbox{ if }q_z>\pzc\cr
-1&\mbox{ if }q_z<-\pzc\cr
1 & \mbox{ if }|q_z|<\pzc\mbox{ and }S_{nq}>\Sc\cr
-1 & \mbox{ if }|q_z|<\pzc\mbox{ and }S_{nq}<-\Sc\cr
0 & \mbox{ otherwise}
\end{array}
\right.
\hspace{0.5cm}
\mbox{and}
\hspace{0.5cm}
\omegasq=\omegas
\left\{
\begin{array}{ll}
1&\mbox{ if }p_z>\pzc\cr
-1&\mbox{ if }p_z<-\pzc\cr
1 & \mbox{ if }|p_z|<\pzc\mbox{ and }S_{np}>\Sc\cr
-1 & \mbox{ if }|p_z|<\pzc\mbox{ and }S_{np}<-\Sc\cr
0 & \mbox{ otherwise}
\end{array}
\right.\,.
\label{eq:optimal_four_signals}
\end{align}
We did not find any strategy performing better than Eq.~(\ref{eq:optimal_four_signals}) using reinforcement learning.
We remark that a similar strategy was found in Fig.~3(d) in Ref.~\cite{Qiu2021} for the signals $p_z$ and $S_{np}$ in two-dimensional frozen flows.
But in that case the strain signal $S_{np}$ was prioritized in the optimal strategy, and the $S_{np}$ signal also dominated over $p_z$ for the case of a single signal.
The explanation for this apparent contradiction is that swimmers that follow the strategy (\ref{eq:optimal}) in two-dimensional frozen flows for the parameters considered in Ref.~\cite{Qiu2021}, tend to end up at stable fixed points or limit cycles, prohibiting them from upward migration unless a significant amount of Brownian noise is added.
This trapping is similar to gyrotactic trapping observed for naive swimmers without steering in shear flows~\cite{Durham2009,Santamaria2014,Cencini2019} and vortical flows~\cite{Durham2011}.
In chaotic frozen 3D flows, or if the flow fluctuates quickly enough due to turbulence, this trapping goes away and it is then beneficial to adopt the strategies for active gyrotactic stability considered here.

Fig.~\ref{fig:velocity_full} compares the robustness of the found strategies to changes in the flow for a range of $\tauK$ with $\urms$ = \SI{1}{\milli\metre\per\s} (upper row) and $\urms$ = \SI{10}{\milli\metre\per\s} (lower row).
The case $\urms$ = \SI{1}{\milli\metre\per\s} can for example be obtained in a seasonal thermocline, where $\tauK$ $\sim$ \SIrange{3}{30}{\s} and the corresponding Taylor scale Reynolds number is relatively small, ${\rm Re}$ $\sim$ \numrange{7}{70}~\cite{Yamazaki1996}.
The case $\urms$ = \SI{10}{\milli\metre\per\s} can be obtained in more turbulent environments, for example a fjord with $\tauK$ $\sim$ \SIrange{0.3}{1}{\s} and ${\rm Re}$ $\sim$ \numrange{70}{200}~\cite{Gargett1984,Yamazaki1996}.
The data in Fig.~\ref{fig:velocity_full} is generated using the statistical model with simulation units such that $\urms$ and $\tauK$ take their desired values in comparison to the parameters in Table~\ref{tab:parameters}.
The left column shows the average vertical velocity component, $\langle v_z\rangle$.
Using Eq.~(\ref{eq:eqm_v}) this can be decomposed as
\begin{align}
\langle v_z\rangle=\langle u_z\rangle+\langle v^{(\rm g)}_z\rangle+\vs\langle n_z\rangle\,.
\label{eq:vdecomposition}
\end{align}
Fig.~\ref{fig:velocity_full} shows the two contributions $\vs\langle n_z\rangle$ (middle column) and $\langle u_z\rangle$ (right column) to $\langle v_z\rangle$, normalized by the vertical velocity in a quiescent flow, Eq. (\ref{eq:vz0}).
We do not show the contribution $\langle v^{(\rm g)}_z\rangle/\langle v_z\rangle_0$ because it is approximately constant and equal to $-0.1$.

First, we compare the planar strategy using $\oomegas$ from Eq.~(\ref{eq:optimal}) [red,$\circ$] to the naive strategy using $\oomegas=0$ [green,$\Box$]. Fig.~\ref{fig:velocity_full}(a,d) shows that the planar strategy has larger average upward velocity $\langle v_z\rangle$ for all considered flow parameters, only being close for very weak flow gradients, $\tauK\sim$ \SI{10}{\s}.
While the naive strategy has small or even negative values of $\langle v_z\rangle$, the planar strategy is of the order of the vertical velocity in quiescent flows, $\langle v_z\rangle_0$, for all considered parameter values, except for the case of strong flow velocity $\urms$ = \SI{10}{\milli\metre\per\s} and strong flow gradients, $\tauK\sim$ \SI{0.1}{\s}.
The explanation to the difference in velocity between the two strategies is the additional alignment $\langle n_z\rangle$ of the planar strategy due to the active gyrotactic stability, Fig.~\ref{fig:velocity_full}(b,e).
Even though the planar strategy obtains a slightly more negative sampling of the vertical flow component, $\langle u_z\rangle$, than the naive case, see Fig.~\ref{fig:velocity_full}(c,f), this difference is negligible compared to the difference in the contribution to the velocity due to alignment, $v_{\rm s}\langle n_z\rangle$.

Next, we consider the full 3D model using two signals (\ref{eq:optimal_two_signals}) [blue,$\Diamond$].
In this model the alignment reaches a plateau, close to perfect alignment, $\langle n_z\rangle\approx 1$, unless the flow gradients are strong.
The mean upward velocity shows the same trend, but with a plateau somewhat below $\langle v_z\rangle_0$ due to the negative contribution from sampling of the vertical flow component, $\langle u_z\rangle<0$.
Finally, for the full 3D model using four signals (\ref{eq:optimal_four_signals}) [$\vartriangle$,magenta], the alignment $\langle n_z\rangle$ is approximately the same as for the case of two signals.
But the refined response using the strain components, leads to a sampling of the vertical flow component $\langle u_z\rangle$ that is less negative and even positive for some parameter values. The net result is a larger positive vertical velocity than the strategy with two signals, allowing vertical migration that is nearly as efficient as in the quiescent flow for some parameters.

Simulation results with $\urms$ = \SI{100}{\milli\metre\per\s} are identical to the data with $\urms$ = \SI{10}{\milli\metre\per\s} within numerical precision (not shown).
In this limit, the active mechanism only starts failing when the flow gradients $\sim 1/\tauK$ are much larger than $\omegas$.
In contrast, the naive strategy fails when the flow gradients are much larger than $1/(2B)$, which is one order of magnitude smaller than $\omegas$ for the parameters from Table~\ref{tab:parameters}. We remark that these parameters are a typical sample. Specific plankton species in nature could have other values of $\omegas$ or $1/(2B)$, adjusting the relative importance of the active or passive reorientation mechanism compared to the flow gradients.
For values of $\urms$ much smaller than \SI{1}{\milli\metre\per\s}, the average alignment approaches unity for all the cases in the displayed range of $\tauK$ (not shown).
In conclusion, the found strategies are robust, showing good performance for a large range of flow parameters.

\section{Conclusions}
\label{sec:conclusions}

We used reinforcement learning to find robust and efficient strategies for vertical migration of microswimmers.
In the simplest case, the swimmer has only access to a single signal $\Delta u_p$, or equivalently $p_z$, which allows for detection of the angle between the orientation of the swimmer (its $\ve p$-axis) and gravity.
By actively rotating in the appropriate direction around its $\ve q$-axis, the swimmer exhibits an active gyrotactic contribution to the angular velocity which is one order of magnitude larger than the passive gyrotactic contribution for typical copepods.
The active contribution results in increased upward alignment, leading to vertical migration of the order of that in quiescent flows unless turbulence is very vigorous.
If the swimmer in addition is able to sense $\Delta u_q$, it can, up to the resolution set by signal thresholds, measure its orientation relative to gravity leading to yet stronger upward alignment and vertical migration velocity.
By using the strain components $S_{np}$ and $S_{nq}$, the swimmer can further increase its migration velocity by exploiting up-welling regions of the flow~\cite{Qiu2021}.
Rotating in the opposite direction allows the swimmer to, even though being bottom heavy, align with gravity, providing an efficient means for downward swimming.

In the passive gyrotactic mechanism, the shape and mass distribution are important.
Using active gyrotactic stability, these parameters are not as important, even spherical swimmers with homogeneous mass distribution ($\Lambda=0$ and $B=\infty$) show the same degree of vertical alignment in turbulent flows. For the parameters in Table~\ref{tab:parameters}, the passive gyrotactic mechanism only becomes relevant in close to quiescent flows, where it allows for better upward alignment than what can be obtained by a single signal $\Delta u_p$.
The active mechanism is robust and efficient over a large range of the flow parameters $\tauK$ and $\urms$. Simulations of the statistical model and DNS give similar results, both when finding the same types of optimal strategies when training in frozen flows and in the evaluation using time-fluctuating flows.

What bearing do our results have for understanding the behavior of plankton in the turbulent ocean?
It is believed that light is important for daily migration of plankton.
Strong light determines the swimming direction of many plankton species~\cite{Ringelberg2009}.
We propose that the mechanism discovered here may serve as a means for efficient vertical migration of plankton species at low light intensities, and may serve as a complementary guide also in the presence of light.
Further experiments are needed to verify whether plankton in nature have evolved to exploit our proposed mechanism.

\begin{acknowledgments}
KG acknowledges support from Vetenskapsrådet, Grant No. 2018-03974 and by a grant from the Knut and Alice Wallenberg Foundation, grant no. 2019.0079.
JQ and LZ were supported by the National Natural Science Foundation of China (Grant No. 11702158) and the Institute for Guo Qiang of Tsinghua University (Grant No. 2019GQG1012). LZ was supported by a collaboration grant from the joint China-Sweden mobility programme (NSFC-STINT) [grant numbers 11911530141 (NSFC), CH2018-7737 (STINT)].
Statistical model simulations were performed on resources provided by the Swedish National Infrastructure for Computing (SNIC).
\end{acknowledgments}

\appendix

\section{Reinforcement learning training parameters}
\label{appendix:learning_parameters}
We train the swimmer using reinforcement learning as described in Section~\ref{sec:model}.
The training parameters are summarized in Table~\ref{tab:learning_parameters} for the cases where states consist of a single signal, two signals, or four signals.
We train in frozen flows obtained from simulations of Eq.~(\ref{eq:NS}) in the DNS and the flow (\ref{eq:SM_u}) with Gaussian statistics (\ref{eq:SM_A}) in the statistical model.
For both flows, units are scaled such that $\urms$ = \SI{6.7}{\milli\metre\per\s} and $\tauK$ = \SI{1}{\s}.
The total number of episodes in a training session were adjusted to reach convergence to approximately optimal strategies. We used $\sim 1000$ episodes for the DNS and $3000$ episodes for the statistical model.

\begin{table}[]
	\centering
    \setlength{\tabcolsep}{6pt}
    \setlength\extrarowheight{0pt}
	\begin{tabular}{lcccccccc}
	& $\gamma$ & $\alpha_0$ & $\varepsilon_0$ & $\Ealpha$ & $\Eeps$ & $\Teq$ & $\Ttrain$ & $\Tmeas$ \\
	  \hline
DNS, 1 signal & 0.99 & 0.01 & 0.005 & 500 & 800 & \SI{50}{\s} & \SI{88}{\s} & \SI{0.0088}{\s}\\
DNS, 2 signals & " & 0.02 & " & " & " & " & " & " \\
DNS, 4 signals & " & 0.15 & 0.001 & " & " & " & " & " \\
Statistical model, 1 signal & " & 0.01 & 0.005 & " & " & \SI{112}{\s} & \SI{447}{\s} & N/A\\
Statistical model, 2 signals & " & 0.02 & " & " & " & " & " & " \\
Statistical model, 4 signals & " & 0.04 & 0.02 & 1000 & 2000 & " & " & " \\
	\end{tabular}
\caption{\label{tab:learning_parameters}Values of the training parameters in our reinforcement learning.}
\end{table}

\section{Calculation of alignment in absence of flow and passive gyrotactic torque}
\label{appendix:alignment}

To derive Eq.~(\ref{eq:nz_planar}), we average the long-time limit of $n_z$ obtained by following the strategy~(\ref{eq:optimal}),
\begin{align}
n_z\to \left\{
\begin{array}{ll}
\sqrt{1-[\pzc]^2-q_{z,0}^2} & \mbox{if }|p_{z,0}|>\pzc\cr
n_{z,0} & \mbox{otherwise}
\end{array}
\right.\,,
\label{eq:nz_limit}
\end{align}
over uniformly distributed initial angles.
To this end, we parameterize the coordinate system $\ve n$, $\ve p$ and $\ve q$ using three angles ($0<\theta<\pi$, $-\pi<\varphi<\pi$ and $-\pi<\alpha<\pi$) as follows
\begin{align}
\label{eq:coordinate system}
\ve n=
\begin{pmatrix}
\sin\theta\cos\varphi\cr
\sin\theta\sin\varphi\cr
\cos\theta
\end{pmatrix}
\,,\hspace{0.3cm}
\ve p=
\cos\alpha\begin{pmatrix}
\sin\varphi\cr
-\cos\varphi\cr
0
\end{pmatrix}
+\sin\alpha\,\ve n\cross\begin{pmatrix}
\sin\varphi\cr
-\cos\varphi\cr
0
\end{pmatrix}
\hspace{0.3cm}\mbox{and}\hspace{0.3cm}
\ve q=\ve n\cross\ve p\,.
\end{align}
Here the directions $\ve p$ and $\ve q$ are orthogonal to $\ve n$, chosen such that $p_z=0$ when $\alpha=0$ and $\ve q$ is chosen to form a right-handed coordinate system.
The angles of uniformly distributed directions are distributed according to $P(\theta,\varphi,\alpha)=\sin\theta/(8\pi^2)$.
Integrating $\varphi$ away and changing coordinates to $p_z=-\sin\alpha\sin\theta$ and $q_z=-\cos\alpha\sin\theta$ gives the joint distribution $P(p_z,q_z)=1/(2\pi\sqrt{1-p_z^2-q_z^2})$ of $p_z$ and $q_z$ with $p_z^2+q_z^2\le 1$.
Using this distribution to average the limiting value (\ref{eq:nz_limit}) for the case $|p_{z,0}|>\pzc$ with $0\le\pzc\le 1$  over the initial orientation gives (the case $|p_{z,0}|\le\pzc$ gives a zero contribution due to equal probability of $n_{z,0}$ taking either sign)
\begin{align}
\langle n_z\rangle&=\int_0^{\sqrt{1-[\pzc]^2}}{\rm d}q_{z,0}\int_{\pzc}^{\sqrt{1-q_{z,0}^2}}{\rm d}p_{z,0} 4P(p_{z,0},q_{z,0})\sqrt{1-[\pzc]^2-q_{z,0}^2}\nonumber\\
&=\frac{2}{\pi}\int_0^{\sqrt{1-[\pzc]^2}}{\rm d}q_{z,0}\sqrt{1-[\pzc]^2-q_{z,0}^2}\;{\rm atan}\Big[\frac{1}{\pzc}\sqrt{1-[\pzc]^2-q_{z,0}^2}\Big]\,.
\end{align}
Here we used symmetry of the integrand to only consider positive values of $p_{z,0}$ and $q_{z,0}$, and the limits of integration are obtained from the conditions $|p_{z,0}|>\pzc$ and $p_{z,0}^2+q_{z,0}^2\le 1$.
The remaining integral can be represented using a generalized hypergeometric function, giving Eq.~(\ref{eq:nz_planar}).

\begin{figure}
\includegraphics[width=5.5cm]{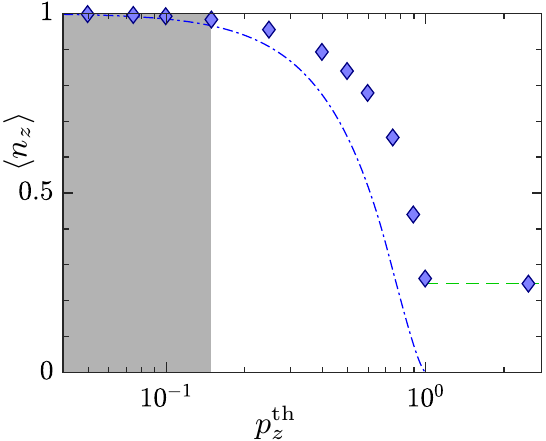}
\caption{
\label{fig:alignment_full}
Same as Fig.~\ref{fig:velocity}(a), but for the full 3D model.
Symbols show numerical results following the strategy in Eq.~(\ref{eq:optimal_two_signals}).
The dash-dotted line shows the theory: Eq. (\ref{eq:alignment_full}) for $\pzc<1/\sqrt{2}$ and two times the value in Eq.~(\ref{eq:nz_planar}) if $1/\sqrt{2}<\pzc\le 1$.
}
\end{figure}

For the case of the strategy using two signals, Eq.~(\ref{eq:optimal_two_signals}), the long-time limit of $n_z$ becomes
\begin{align}
n_z\to \left\{
\begin{array}{ll}
\sqrt{1-[\pzc]^2-q_{z,0}^2} & \mbox{if }|p_{z,0}|>\pzc\mbox{ and }|q_{z,0}|\le\pzc\cr
\sqrt{1-[\pzc]^2-p_{z,0}^2} & \mbox{if }|p_{z,0}|\le\pzc\mbox{ and }|q_{z,0}|>\pzc\cr
\sqrt{1-2[\pzc]^2} & \mbox{if }|p_{z,0}|>\pzc\mbox{ and }|q_{z,0}|>\pzc\cr
n_{z,0} & \mbox{otherwise}
\end{array}
\right.\,.
\label{eq:nz_limit_two_signals}
\end{align}
Averaging this expression over uniformly distributed angles as above, gives that $\langle n_z\rangle$ is two times the value in Eq.~(\ref{eq:nz_planar}) if $1/\sqrt{2}<\pzc\le 1$, and that
\begin{align}
\begin{split}
\langle n_z\rangle
&=\sqrt{1-2[\pzc]^2}\bigg(1-\frac{4\pzc}{\pi}{\rm acos}\Big[\frac{\pzc}{\sqrt{1-[\pzc]^2}}\Big]-\frac{1}{\pi}{\rm acos}\Big[\frac{1-2[\pzc]^2-[\pzc]^4}{(1-[\pzc]^2)^2}\Big]\bigg)
\\
&\hspace{1cm}+\frac{4}{\pi}\int_{0}^{\pzc}{\rm d}p_{z,0}\sqrt{1-p_{z,0}^2-[\pzc]^2}\,{\rm acos}\Big[\frac{\pzc}{\sqrt{1-p_{z,0}^2}}\Big]
\end{split}
\label{eq:alignment_full}
\end{align}
if $\pzc<1/\sqrt{2}$.
Eq.~(\ref{eq:alignment_full}) is compared to numerical simulations in Fig.~\ref{fig:alignment_full}.
The theory agrees qualitatively with the numerical simulations, similar to the theory in the planar model in Fig.~\ref{fig:velocity}(a).
We did not find a representation of the last integral in Eq.~(\ref{eq:alignment_full}) in terms of standard functions, but it is straightforward to evaluate the term in a series expansion for small $\pzc$, suitable for the interval $0<\pzc<1/\sqrt{2}$. The first few terms in the vertical alignment becomes
\begin{align}
\langle n_z\rangle\sim 1-\Big(1+\frac{2}{\pi}\Big)\left[\pzc\right]^2+\frac{2}{3}\left[\pzc\right]^3-\frac{1}{2}\left[\pzc\right]^4+\frac{8}{15}\left[\pzc\right]^5\,.
\end{align}

\bibliographystyle{ieeetr}

\end{document}